\documentclass[superscriptaddress, 12pt, aps, prl,  preprint, longbibliography]{revtex4-1}
\usepackage{graphicx}
\usepackage{epstopdf}
\usepackage{color}
\usepackage{amsmath, amssymb}

\begin{document}

\title{Van der Waals interactions in supercritical water under Earth's mantle conditions}
\author{Jiajia Huang}
\affiliation{Department of Physics, The Hong Kong University of Science and Technology, Clear Water Bay, Hong Kong, P. R. China.}
\author{Rui Hou}
\affiliation{Department of Physics, The Hong Kong University of Science and Technology, Clear Water Bay, Hong Kong, P. R. China.}
\author{Ding Pan}
\email{dingpan@ust.hk}
\affiliation{Department of Physics, The Hong Kong University of Science and Technology, Clear Water Bay, Hong Kong, P. R. China.}
\affiliation{Department of Chemistry, The Hong Kong University of Science and Technology, Clear Water Bay, Hong Kong, P. R. China}

\date{\today}

\begin{abstract}
  The properties of water under high pressure and high temperature (HP-HT) are critical in multiple geochemical processes in deep Earth. Ab initio molecular dynamics (AIMD) is a promising approach to study water under extreme conditions without any empirical parameters. However, the accuracy of AIMD simulations is determined by the exchange-correlation (XC) functional including the dispersion correction used in density functional theory calculations. While van der Waals (vdW) interactions are well known to be critically important for water under ambient conditions, the influence of the dispersion correction on the properties of HP-HT water as found in Earth's mantle remains largely unexplored. To address this, we carried out AIMD simulations for supercritical water at 1, 5, and 10 GPa, and 1000 K. We compared PBE, PBE-D3, RPBE-D3, and SCAN functionals, where D3 means Grimme's D3  
  dispersion correction.  We compared the structural, diffusion, and vibrational properties of water as computed with these XC functionals. Overall, the discrepancies between the functionals are reduced under extreme P–T conditions relative to ambient conditions. PBE and PBE-D3 exhibit higher proton-transfer rates than RPBE-D3 and SCAN, suggesting that while vdW interactions do not significantly affect the water structure under extreme conditions, the oxygen–hydrogen bond strength does influence proton transfer. Our results provide molecular-level insight into water in Earth's mantle and offer valuable guidance for selecting appropriate XC functionals in AIMD simulations of aqueous solutions under extreme conditions.

\end{abstract}

\maketitle

Water, essential for life at Earth’s surface, also plays a foundational role deep within our planet, stored throughout the crust, mantle, and potentially the core \cite{mao2017water,he2026absence}. For decades, scientific understanding held that Earth’s mantle water existed primarily as trace hydrogen or hydroxyl ions locked within mantle minerals. A paradigm shift is now underway. Recent geochemical evidence compellingly suggests that a free, mobile aqueous fluid phase can exist within the upper mantle and may persist into the transition zone and beyond \cite{tschauner2018ice}.  
In the Earth's upper mantle, where pressures can exceed 10 GPa and temperatures can reach up to 1700 K \cite{thompson1992WaterEarthUpper}, water serves as a carrier for carbon and minerals and facilitates important recycling processes 
\cite{manning2013chemistry, manning2020subduction, ohtani2020role, economon2026complexity, stolte2026formation}.

Under such high-pressure and high-temperature (HP-HT) conditions, the properties of water and aqueous fluids differ significantly from those at ambient conditions.
Experimentally studying water under extreme HP-HT conditions is challenging due to the inherent difficulties associated with experimental techniques \cite{forster2024generating}. The experimental methodologies that are effective at ambient pressure and temperature conditions fall short under the extreme conditions. Consequently, alternative computational approaches, such as first-principles molecular dynamics (AIMD) simulations, have emerged as accurate and parameter-free tools \cite{schweglerWaterPressure2000, pan2013DielectricPropertiesWater, rozsa2018InitioSpectroscopyIonic, yu2026predicting}. 

Nevertheless, the accuracy of AIMD simulations relies on the choice of exchange-correlation functional, including the incorporation of van der Waals (vdW) interactions within density functional theory (DFT) calculations \cite{gillan2016PerspectiveHowGood}. 
The commonly used generalized gradient approximation (GGA) in DFT for liquid water, such as Perdew-Burke-Ernzerhof (PBE) functional\cite{PhysRevLett.77.3865}, leads to an overly structured and slow-diffusing water at ambient conditions \cite{gillan2016PerspectiveHowGood}. 
Despite attempts to improve the PBE functional by modifying its exchange energy part, as seen in revPBE\cite{PhysRevLett.77.3865,zhang1998comment}, RPBE\cite{PhysRevB.59.7413} and hybrid functionals with Hartree-Fock, these modifications are still unable to achieve satisfactory performance due to the absence of non-local correlations associated with vdW interactions.

The vdW interactions can be viewed as attractive forces that arise from the correlation of instantaneous, quantum-mechanical fluctuations in the electron densities of atoms or molecules, which induce dipoles in one another. At intermediate distances, this attractive interaction energy decays as $-1/r^6$ with the interatomic distance r. 
Recently, the inclusion of such dispersion interactions in standard GGA functionals\cite{biswasInterstitialVoidsResultant2018,chenInitioTheoryModeling2017c,morawietzHowVanWaals2016c,bankuraStructureDynamicsSpectral2014b,forster-tonigoldDispersionCorrectedRPBE2014b} has led to a much better match with experimental observations, including the density, structure, and dynamics for ambient water.
The methods for including vdW interactions can be categorized into three types \cite{grimmeDensityFunctionalTheory2011, klimes2012PerspectiveAdvancesChallenges,grimme2016DispersionCorrectedMeanFieldElectronic}: 
(i) methods relying solely on electron density-based information \cite{PhysRevLett.92.246401}, (ii) methods employing artificial one-electron potentials (1ePOT) \cite{PhysRevLett.93.153004, 10.1063/1.3651474}, and (iii) methods based on semiclassical atom-pairwise potentials \cite{grimme2004accurate, grimme2006semiempirical}. Among these, Grimme's D3 type correction\cite{grimmeConsistentAccurateInitio2010,grimme2011EffectDampingFunction}, an atom pairwise $C_6$ based potential approach, stands out for its promising balance between low computational cost and satisfactory performance for ambient water\cite{biswasInterstitialVoidsResultant2018,dodia2019StructureDynamicsWater,lin2012StructureDynamicsLiquida}. In this correction, the vdW interaction energy $E^{\text{DFT-D3}}$ is simply added to the standard Kohn-Sham DFT result for each atom pair A and B in the form of:
\begin{equation}
  E^{\text {DFT-D3 }}=-\sum_{n=6,8} \sum_{A<B} s_n \frac{C_n^{A B}}{\left(R_{A B}\right)^n} f_{\text {damp,n}}\left(R_{A B}\right)+E^{(3)}
  \end{equation}
where $C_n^{AB}$ is the environment-dependent coefficient for atoms A and B, $R_{A B}$ is the distance between A and B, $s_n$ is a scaling factor typically used to adjust the correction to the repulsive behavior of the chosen XC, $f_{\text {damp,n}}\left(R_{A B}\right)$ is the short-range damping function, and $E^{(3)}$ is a three-body interaction term. 

The effect of vdW interactions should be carefully examined when simulating water systems.  However, under HP-HT conditions, bare GGAs remain prevalent due to their more accurate representation of various experimental quantities than at ambient conditions, such as the equation of state\cite{PhysRevLett.108.105502, pan2013DielectricPropertiesWater}, the dielectric constant\cite{pan2013DielectricPropertiesWater, pan2014RefractiveIndexElectronic}, and the ionic conductivity\cite{rozsa2018InitioSpectroscopyIonic}. Despite this, a comprehensive assessment of the impact of vdW interactions on HP-HT water is still needed.

Here, we present AIMD simulations of water under pressure at 1000K, focusing on the structural, diffusion and dynamical properties of water before molecular dissociation occurs\cite{rozsa2018InitioSpectroscopyIonic}. 
We use GGA level functionals, i.e., the PBE and RPBE functionals, comparing results with and without D3 correction. Additionally, we included a higher level meta-GGA XC functional, i.e., the Strongly Constrained and Appropriately Normed Semilocal Density (SCAN) functional \cite{PhysRevLett.115.036402}, which is meticulously parameterized to address the intermediate and short-range regions where non-local correlation is expected to contribute most to bonding. We also performed AIMD simulations at ambient conditions to enable a direct comparison with available experimental data and to assess the accuracy of the theoretical model.

\begin{figure}[htbp]
  \centering
  \includegraphics[width=1\textwidth]{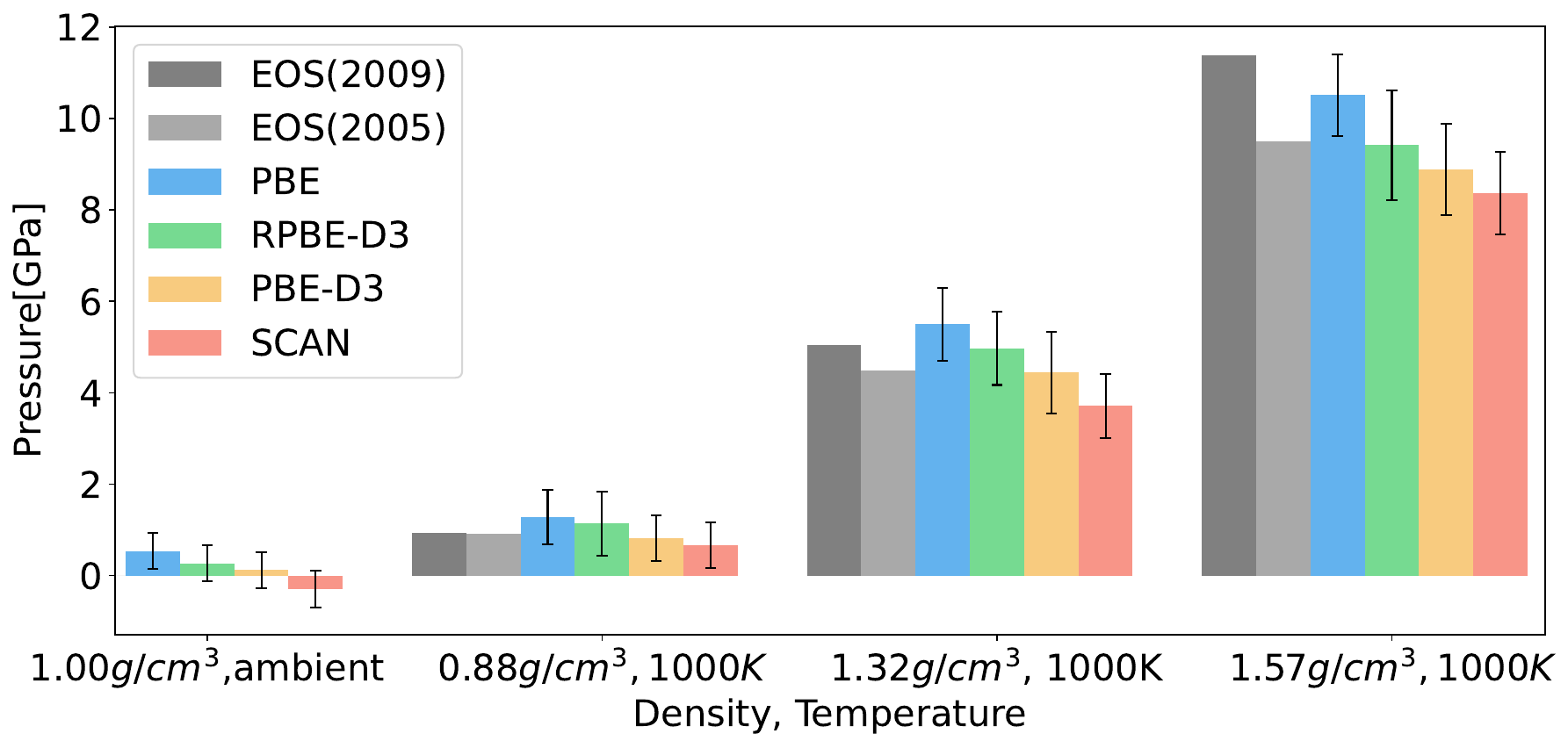}
  \caption{Pressure(GPa) calculated based on different XC functionals at ambient and HP-HT conditions from the analysis of the stress tensor. The results of SCAN(330K) and PBE(400K) for ambient water are from Ref.\cite{lacount2019ensemble}\cite{dawson2018equilibration} }
  \label{pressure-fig}
 \end{figure}

We first compared the equations of state (EOS) of water obtained by different XC functionals. The computed averages of the diagonal elements of the stress tensor of simulation boxes are shown in Fig. \ref{pressure-fig}. At ambient conditions, all functionals produce pressures significantly away from the expected 101 kPa, while the large pressure fluctuations are partly attributable to the limited size of the simulation cell. Given that the ideal mean pressure should be around 101 kPa, the PBE-D3, RPBE-D3, and SCAN functionals describe the liquid density more accurately than PBE at ambient conditions.

With increasing temperature to 1000 K and pressure to a few GPa, experimental measurements become more challenging and experimental data usually have larger error bars
\cite{larrieu1997MeasurementsPressurevolumetemperatureProperties, brodholt1994MeasurementsPVTProperties, withers2000NewMethodDetermining}. The scarcity of experimental PVT data in these relevant regions has led numerous studies to report the equation of state (EOS) of water by combining extrapolations of existing experimental and MD simulation data \cite{wagner2002IAPWSFormulation1995, abramson2004EquationStateWaterb, zhang2005PredictionPVTProperties, liu2020ThermodynamicModelsH2O, liu2019PredictionH2OPVT}. 
Here we compared our calculated pressures with two EOS models commonly applied to HP-HT water: EOS(2005) \cite{zhang2005PredictionPVTProperties} and EOS(2009) \cite{zhang2009ModelFluidEarth}. 
The pressures predicted by SCAN are lowest, followed by PBE-D3 and RPBE-D3, and PBE has the highest pressure for every given density and temperature.
Unlike at ambient conditions, 
PBE without the vdW corrections also gives reasonable pressures between 1 and $\sim$10 GPa.

  \begin{figure} 
    \centering
    \includegraphics[width=1\textwidth]{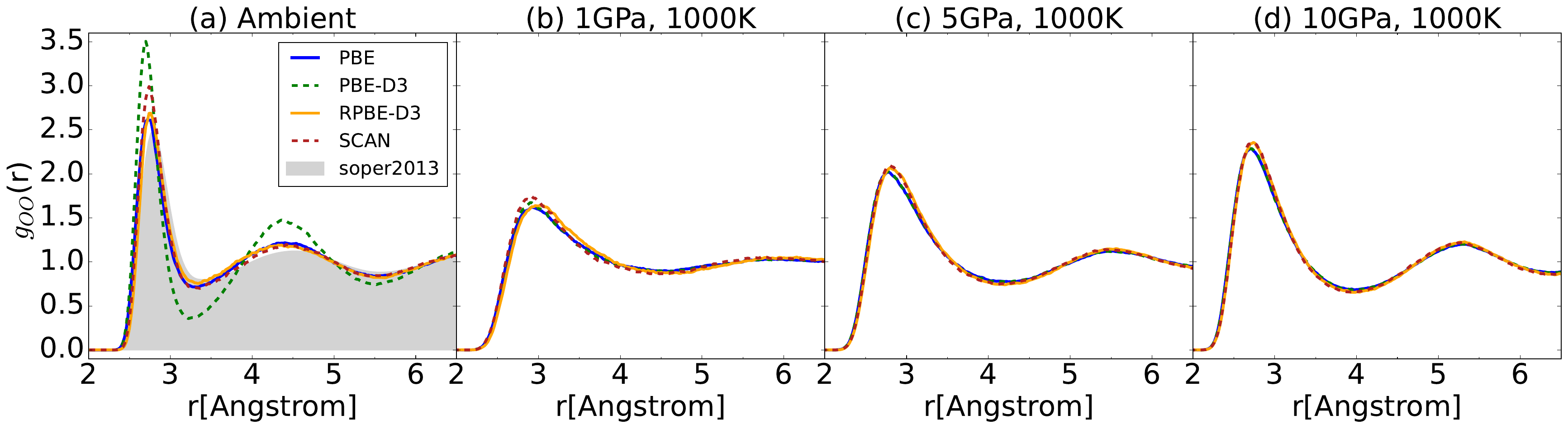}
    \caption{O–O RDFs of water at ambient temperature and 1000 K with a pressure of 1 GPa, 5 GPa and 10 GPa based on the PBE, PBE-D3, RPBE-D3 and SCAN approximations. The results of SCAN(330K) and PBE(400K) for ambient water are from Ref.\cite{lacount2019ensemble}\cite{dawson2018equilibration}}
    \label{o-o}
  \end{figure}

X-ray and neutron diffraction experiments are commonly used to obtain the radial distribution function (RDF) of water, which can provide important information about the molecular structure of water.
Here, we calculated the RDFs of oxygen atoms in water ($g_{OO}$) at ambient conditions, 
and compared our results with the experimental data at 298 K \cite{soper2013radial} to evaluate the performance of XC functionals.
We found significant discrepancies among the four XC functionals, as shown in Fig. \ref{o-o}(a).  
RPBE-D3 gives the closest RDF to the experimental data. 
The directed hydrogen bonds overestimated by PBE are replaced by the weaker hydrogen bonds described by RPBE and corrected by the addition of non-directed vdW interactions. 
It is reported that the PBE water may freeze at ambient temperature, so the temperature was often increased to about 400 K to simulate ambient water in many previous studies\cite{dawson2018equilibration, schwegler2004AssessmentAccuracyDensity, sit2005StaticDynamicalProperties}.
The RDF obtained by PBE at 400 K is very similar to that from RPBE-D3.
At 330 K, we replaced RPBE by PBE while keeping the D3 correction, and found the water became overstructured. 
It seems adding the D3 dispersion correction to PBE still cannot overcome the overbinding of water molecules \cite{biswasInterstitialVoidsResultant2018},
which is consistent with the previous studies\cite{forster-tonigoldDispersionCorrectedRPBE2014b, lin2012StructureDynamicsLiquida}.
The SCAN functional, which includes a portion of the dispersion force, shows improved performance compared to PBE-D3, but still results in slightly over-structured water at 330 K.

At high pressure and 1000 K, Fig. \ref{o-o}(b, c, d) shows that the first peaks in the RDFs of oxygen atoms in water are broadened compared with those at ambient conditions, indicating that water at HP-HT becomes less structured than at ambient conditions. 
Interestingly, the RDFs obtained from the PBE, PBE-D3, RPBE-D3, and SCAN functionals 
show remarkable similarity, which is very different from their behavior under ambient conditions.
This similarity implies that the vdW corrections have a relatively minor effect on water under HP-HT conditions.
A possible explanation is that, under extreme conditions, electron densities show greater overlap than at ambient conditions, which leads to improved performance of the semilocal approximation in XC functionals.

  \begin{figure} 
    \centering
    \includegraphics[width=0.5\textwidth]{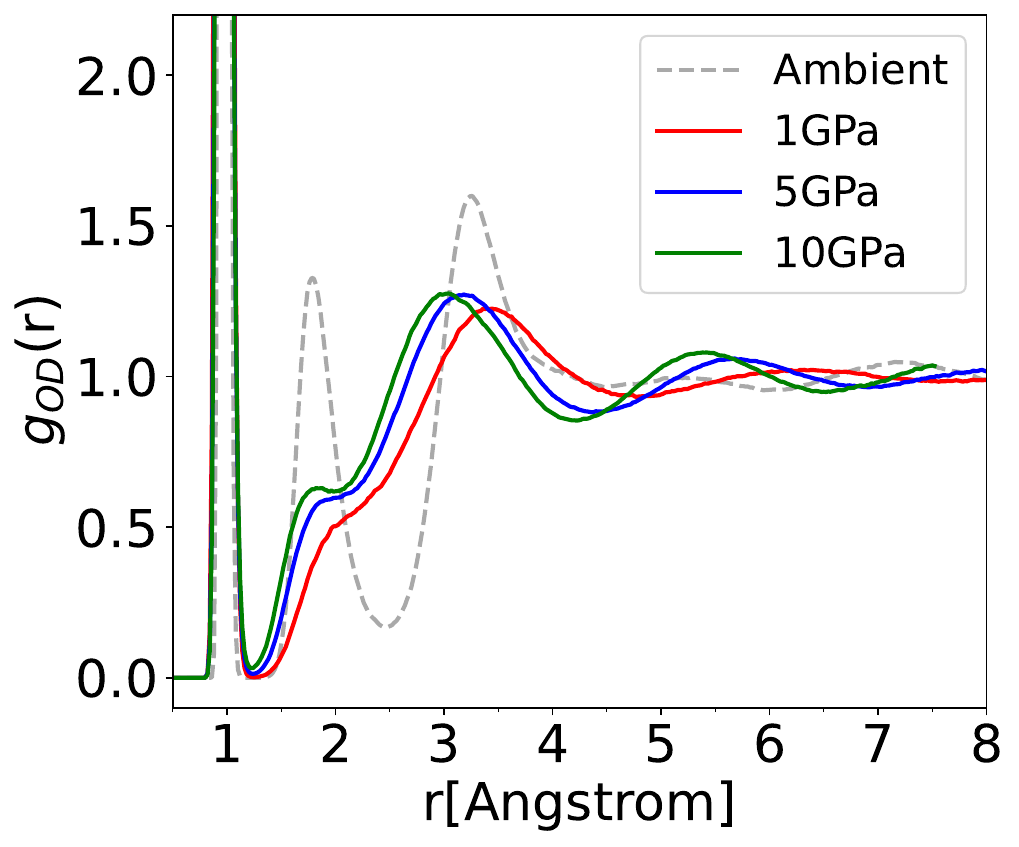}
    \caption{O–D RDFs of water at ambient temperature and 1000 K with a pressure of 1 GPa, 5 GPa and 10 GPa based on the RPBE-D3 approximation.}
    \label{o-h-rpbe-d3}
  \end{figure}

In the following part, we chose the RPBE-D3 functional, which yielded RDFs most similar to experimental results at ambient conditions, for further examination of the pressure effect on the structure of water (Fig. \ref{o-h-rpbe-d3}). 
With increasing pressure from 1 to 10 GPa, the peaks in $g_{OD}$ near 2 and 3~\AA\ shift toward shorter distances, and the number of hydrogen atoms surrounding each water molecule increases. These trends are consistent with previous studies of high-pressure water \cite{ikedaCommunicationsHightemperatureWater2010,schweglerWaterPressure2000}. Compared with ambient conditions, the second peak of $g_{OD}$ near 2~\AA\ is significantly reduced, indicating that the complex hydrogen-bond network breaks down under HP-HT conditions. This disruption may explain the reduced sensitivity of water to vdW corrections under HP-HT conditions.

\begin{figure}[htbp]
  \centering
  \includegraphics[width=0.5\textwidth]{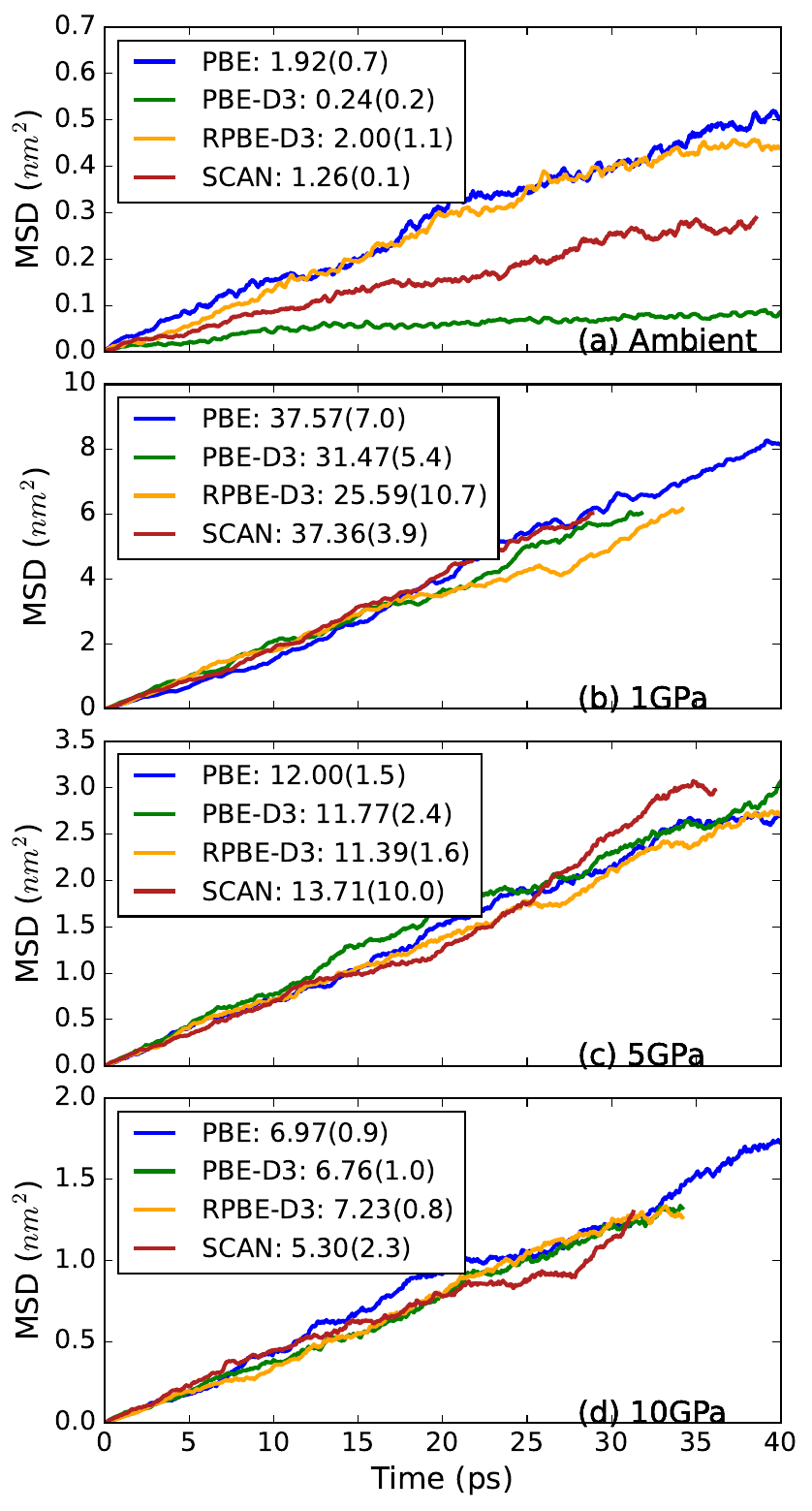}
  \caption{The MSDs of oxygen atoms at ambient conditions and 1000 K with pressure of 1 GPa, 5 GPa and 10 GPa based on the PBE, PBE-D3, RPBE-D3 and SCAN approximations. After the colon is the diffusion constant, unit is 1e-5 $cm^2/s$, with the variance in parentheses. The results of SCAN(330K) and PBE(400K) for ambient water are from Ref.\cite{lacount2019ensemble}\cite{dawson2018equilibration}}
  \label{msd}
\end{figure}

Fig. \ref{msd} shows the mean square displacement (MSD) of oxygen atoms as a function of time, along with the calculated diffusion constant $\sigma$, which is obtained from Einstein's relation by performing a least-squares linear fit to the MSD curve:

\begin{equation}
  \sigma=\frac{1}{6} \lim _{t \rightarrow \infty} \frac{d}{d t}\left\langle[\mathbf{r}(t)-\mathbf{r}(0)]^2\right\rangle
  \end{equation}
where $t$ is the simulation time, $\mathbf{r}$ is the position vector, and $\langle...\rangle$ denotes the average among all the oxygen atoms.

The experimental diffusion constant for heavy water at 298 K is $1.872\times10^{-5}~\text{cm}^2/\text{s}$ \cite{mills1973SelfdiffusionNormalHeavy}. Among the simulations performed at 330 K, RPBE-D3 gives a diffusion constant closest to the experimental value. PBE-D3 yields a very small diffusion constant at 330 K, suggesting glass-like behavior, which is consistent with the overstructured RDF shown in Fig.~\ref{o-o}.

At HP-HT conditions, water remains fluid-like, and the diffusion constant decreases with increasing pressure, aligning with the established phase diagram of water \cite{rozsa2018InitioSpectroscopyIonic}. The variance in diffusion constants among different XC functionals is smaller than under ambient conditions.
A comparison between PBE and PBE-D3 at HP-HT reveals that the inclusion of vdW corrections results in slightly lower diffusional motion. 
A possible explanation is that the rapid reorientation of water molecules at HP-HT, which suppresses molecular diffusion \cite{ikedaCommunicationsHightemperatureWater2010}, is compensated by the inclusion of vdW corrections. It is interesting that this trend differs from ambient conditions, where including vdW corrections in the PBE functional was found to slightly enhance the diffusion of water \cite{lin2012StructureDynamicsLiquida}.

For a 64 molecules supercell, the structural properties are well converged and reproducible. However, studying dynamical properties like diffusivity requires longer simulation times and larger system sizes \cite{kuhne2009StaticDynamicalProperties}. Although it was suggested that one should run multiple independent simulations of computationally manageable durations to produce useful statistical accuracy \cite{pranami2015EstimatingErrorDiffusion,lacount2019ensemble,dawson2018equilibration}, comparisons between different XC functionals in our study are still valid provided that the simulations maintain consistent initial configurations.


We calculated the velocity density of states (VDOS) by the Fourier transformation of the atomic velocity-velocity autocorrelation function: 

\begin{equation}
  VDOS(\omega)= \int_{-\infty}^{\infty}\sum_{n=1}^N \left\langle\mathbf{v}_n(t) \cdot \mathbf{v}_n(0)\right\rangle e^ {-i \omega t} d t,
  \label{vdoseq}
  \end{equation}
where $\omega$ represents the frequency, $\mathbf{v}_n$ is the velocity of the $n$th atom, $N$ is the total number of atoms, $t$ is the correlation time, and $\langle...\rangle$ is the ensemble average in MD simulations. 

\begin{figure}[htbp]
  \centering
  \includegraphics[width=1\textwidth]{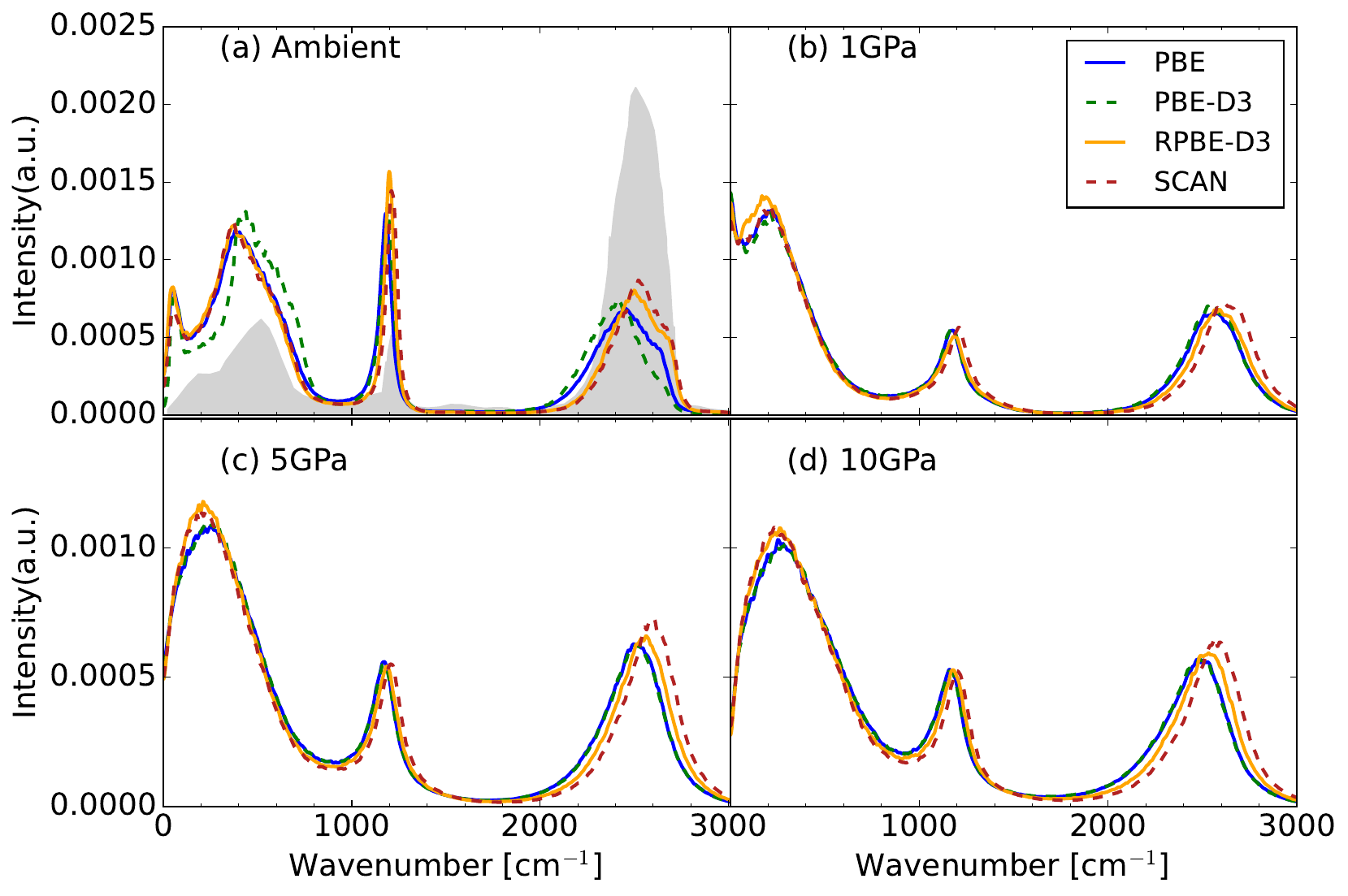}
  \caption{Velocity density of states of water at ambient conditions and 1000K with pressure of 1GPa, 5GPa and 10GPa based on the PBE, PBE-D3, RPBE-D3 and SCAN approximations. The gray region represents the experimental measurement of the IR spectra at 298 K from Ref.\cite{maxIsotopeEffectsLiquid2009}. The results of SCAN(330K) and PBE(400K) for ambient water are from Ref.\cite{lacount2019ensemble}\cite{dawson2018equilibration} }
  \label{vdos}
  \end{figure}

The VDOS describes the frequency distribution of all vibrational modes, irrespective of their optical activity. In many AIMD studies, the VDOS has been used to compare with experimental IR and Raman spectra, which respectively reflect only those modes that induce a change in the system's dipole moment or polarizability \cite{rozsa2018InitioSpectroscopyIonic,zhuInitioMolecularDynamics2002, zhang2011structural, wan2013raman, li2025ab}.
Although the raw VDOS does not reproduce the experimental intensities, the peak positions of the dominant vibrational modes provide a basis for comparison. Accordingly, we compare our computed VDOS at ambient conditions with the experimental IR spectrum measured at 298 K in Fig. \ref{vdos}.

The vibrational modes of heavy water at ambient conditions include the intermolecular $\sim$200 cm$^{-1}$ hindered translations, the $\sim$500 cm$^{-1}$ intermolecular libration, the $\sim$1200 cm$^{-1}$ DOD bending, and the $\sim$2500 cm$^{-1}$ OD stretching mode. These qualitative shapes are well captured by our VDOS results, despite the absence of IR selection rules. At ambient conditions, the SCAN functional presents the highest OD stretching frequency, followed by RPBE-D3, PBE, and PBE-D3. This suggests that SCAN and RPBE-D3 exhibit a more distorted hydrogen bond network, characterized by a larger number of broken intermolecular hydrogen bonds and stronger covalent bonds compared to PBE-D3 and PBE. At HP and 1000K, this trend remains, and we found that SCAN is like RPBE-D3, while the discrepancy between PBE and PBE-D3 is minor, indicating that the vdW correction does not significantly affect OD stretching.

\begin{figure}[htbp]
  \centering
  \includegraphics[width=0.5\textwidth]{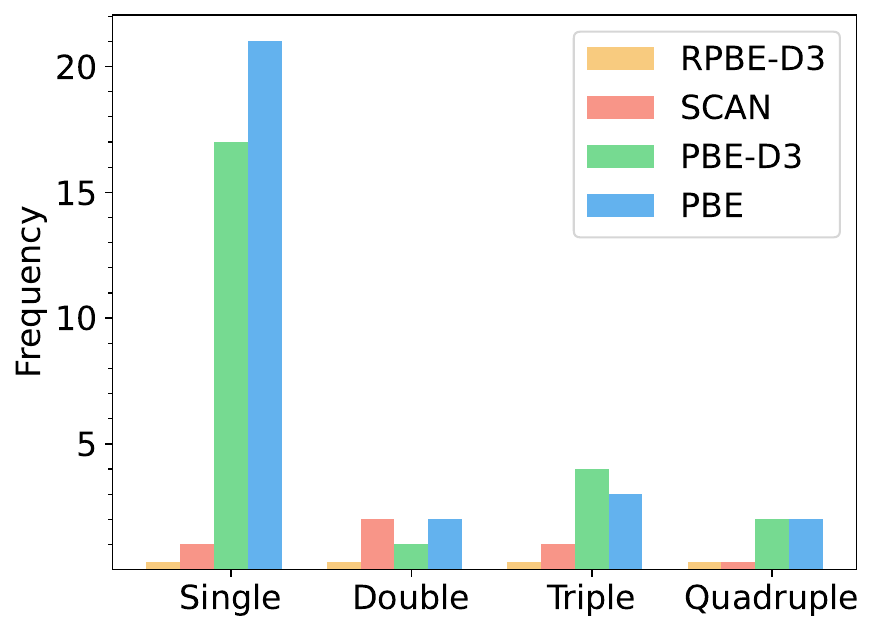}
  \caption{Frequency of proton transfer events with four XC functionals (PBE, PBE-D3, RPBE-D3, SCAN). The jumps of a proton from one oxygen site to another were traced by directly monitoring the proximity of the proton to the nearest oxygen. The frequency was calculated by counting the number of proton transfers of each trajectory at 10 GPa and 1000 K during a time period of 31 ps. Consecutive jumps separated in time by 0.5 ps or less would contribute to multiple proton transfer events. Besides, events in which a proton returned to its original site within 0.5 ps were considered to be rattling fluctuations and were not included in these counts\cite{chenHydroxideDiffusesSlower2018}. The criteria could be adjusted, but the qualitative characteristic would not change. }
  \label{proton}
\end{figure}

At HP-HT conditions, water molecules undergo continuous disruption and reformation of hydrogen-bonded structures. Proton transfer events differ with the choice of functional \cite{chenHydroxideDiffusesSlower2018}. We studied the proton transfer events as shown in Fig. \ref{proton}, in which protons can hop from one water molecule to another in a Grotthuss-like fashion\cite{degrotthuss1806sur}. 
Proton-transfer events occurred most frequently with the PBE and PBE-D3 functionals and were rare with SCAN. Moreover, none were observed with RPBE-D3 within 31 ps.
Our VDOS calculations show that SCAN and RPBE-D3 produce higher OD stretching frequencies than PBE and PBE-D3, which suggests stronger OD bonding and thus greater resistance to OD bond breaking during proton transfer.
It has also been reported that proton transfer in water is a complex process, governed by the collective structural and dynamical characteristics of the hydrogen bond network \cite{hassanali2013proton}.
Because PBE and PBE-D3 give a more structured hydrogen bond network than SCAN and RPBE-D3, the resulting proton hopping probability is consequently enhanced.

In conclusion, our study assessed the performance of PBE, PBE-D3, RPBE-D3, and SCAN functionals in predicting the structural, diffusion, and dynamical properties of water under HP-HT conditions. We observed that various XC functionals exhibit similar RDFs and diffusion constants, suggesting a reduced influence of vdW interactions on the structural and diffusion properties of water under extreme conditions compared to ambient conditions. 
This reduction may be attributed to the substantial disruption of the hydrogen-bond network by HP and HT, where pressure and thermal effects dominate the molecular interactions.
The vibrational density of state calculations show that PBE and PBE-D3 exhibit lower oxygen-hydrogen bending and stretching frequencies, resulting in more frequent proton-transfer events at 1000 K and 10 GPa. These findings enhance our understanding of water's behavior in HP-HT environments and offer valuable insights for future research in geochemistry and materials science.

\section{Methods}

We conducted AIMD simulations in a cubic cell with periodic boundary conditions using the Qbox code, v.1.73.2 \cite{gygi2008ArchitectureQboxScalable} \footnote{http://eslab.ucdavis.edu/software/qbox/}. 
We implemented the van der Waals correction D3\cite{grimme2011EffectDampingFunction} in the Qbox code. 
We used SG15 ONCV norm-conserving pseudopotentials\cite{hamann2013OptimizedNormconservingVanderbilt, schlipf2015OptimizationAlgorithmGeneration} and the plane wave basis sets with a kinetic energy cutoff of 65 Ry, which was increased to 85 Ry for the calculation of pressure. Our simulations were performed in the NVT ensemble using the Bussi-Donadio-Parrinello  thermostat \cite{bussi2007CanonicalSamplingVelocity}, which had a relaxation time of 24.2 fs.
 The XC functionals used here were PBE\cite{PhysRevLett.77.3865}, PBE-D3\cite{PhysRevLett.77.3865, grimme2011EffectDampingFunction}, RPBE-D3\cite{PhysRevB.59.7413, grimme2011EffectDampingFunction}, and SCAN\cite{PhysRevLett.115.036402}. We ran independent AIMD simulations for each functional at 1000 K. 
Our cubic supercell with periodic boundary conditions contained 64 heavy water molecules ($D_{2}O$). 
We used heavy water to enable a large molecular dynamics time step of 0.24 fs, while light water was used to calculate the density.
The simulation box was adjusted to have densities of 0.88, 1.32, and 1.57 $ g/cm^3 $, which align with the findings of a previous study on HP-HT water \cite{pan2013DielectricPropertiesWater}.  These densities correspond approximately to pressures of 1 GPa, 5 GPa, and 10 GPa, respectively.
For ambient conditions, two independent AIMD simulations for PBE-D3 and RPBE-D3 were performed at 330 K at the density of 1.00 $ g/cm^3 $. 
The results for the PBE and SCAN functionals under ambient conditions were obtained from the previous studies\cite{dawson2018equilibration,lacount2019ensemble}, where the simulation at the PBE level
was conducted at an elevated temperature of 400 K to approximate the properties of the liquid at 300 K \cite{dawson2018equilibration}.

\section{Data Availability}

The data that support this study are available upon request from the authors.

\section{Code Availability}
Qbox is a free and open source code available at http://qboxcode.org. Modifications to Qbox and data processing scripts are available upon request from the authors.

\section{Acknowledgements}
This work was supported by the Hong Kong Research Grants Council (RGC) (Projects GRF-16301723, GRF-16306621, GRF-16302423, and GRF-16310225), and National Natural Science Foundation of China/RGC Joint Research Scheme (N\_HKUST664/24).
Part of this work was carried out using computational resources from the National Supercomputer Center in Guangzhou, China, and the X-GPU cluster supported by the RGC Collaborative Research Fund C6021-19EF.

\bibliographystyle{unsrt}
\bibliography{reference.bib}

\end{document}


\title{Supplementary Information for ``Van der Waals interactions in supercritical water under Earth's mantle conditions"}
\author{Jiajia Huang}
\affiliation{Department of Physics, The Hong Kong University of Science and Technology, Clear Water Bay, Hong Kong, P. R. China.}
\author{Rui Hou}
\affiliation{Department of Physics, The Hong Kong University of Science and Technology, Clear Water Bay, Hong Kong, P. R. China.}
\author{Ding Pan}
\email{dingpan@ust.hk}
\affiliation{Department of Physics, The Hong Kong University of Science and Technology, Clear Water Bay, Hong Kong, P. R. China.}
\affiliation{Department of Chemistry, The Hong Kong University of Science and Technology, Clear Water Bay, Hong Kong, P. R. China}

\maketitle

\section{Water dimer}
\begin{figure}[htbp]
  \centering
  \includegraphics[width=0.5\textwidth]{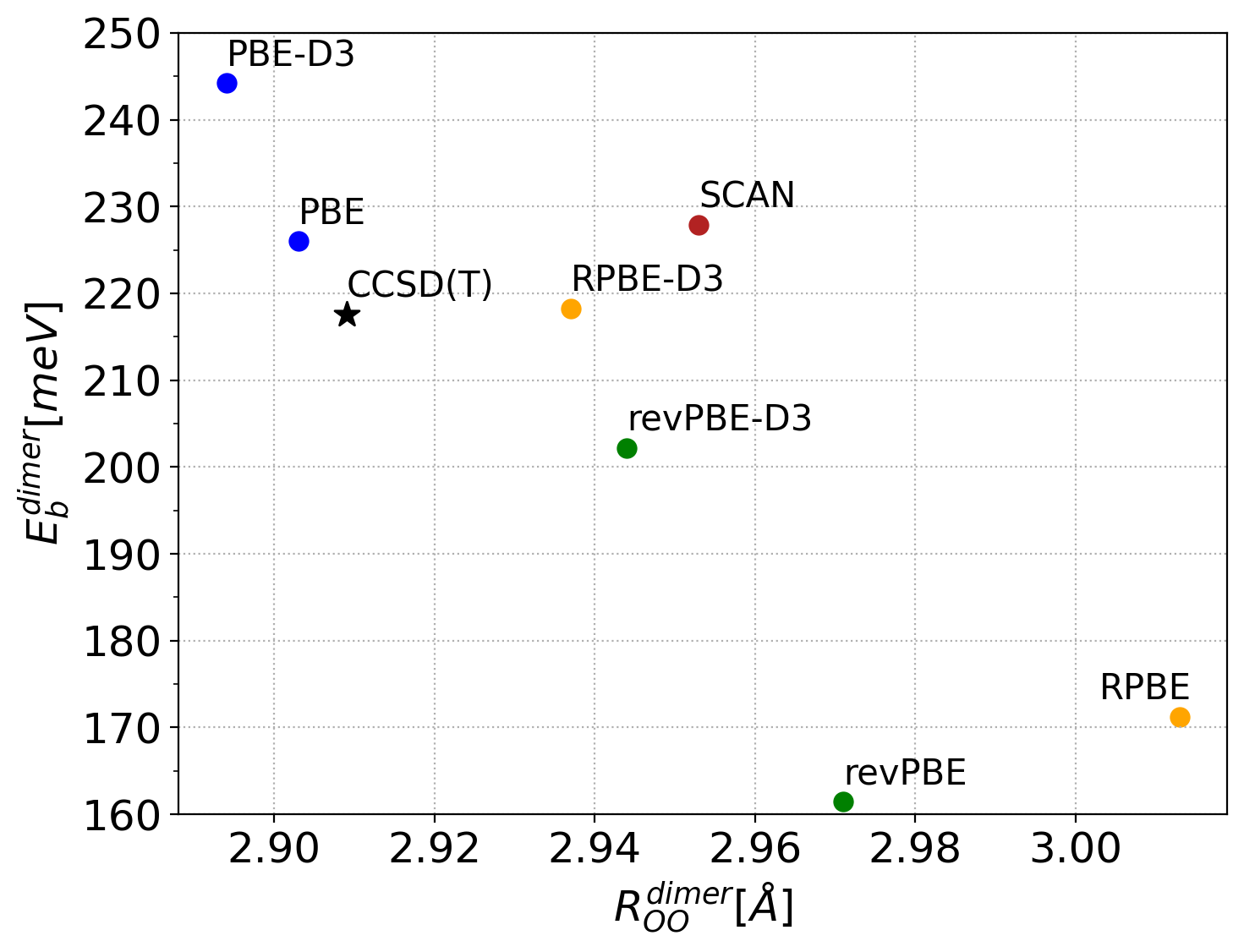}
  \caption{Binding energies $E_{b}^{dimer}$ and equilibrium O–O distances $R_{dist}^{dimer}$ of the $H_{2}O$ dimer in its minimum configuration computed with a variety of semi-local and dispersion-inclusive XC approximations. Benchmark values from CCSD(T) calculations are obtained from ref. \cite{tschumper2002AnchoringWaterDimer}}
  \label{dimer}
 \end{figure}
 
To offer insights into the performance differences observed among the functionals investigated in this study, we show the binding energies ($E_{b}^{dimer}$) and the equilibrium distance between oxygen atoms($R_{dist}^{dimer}$) of water dimer given by various XC functionals and compared to the CCSD(T)\cite{tschumper2002AnchoringWaterDimer} results in Fig. \ref{dimer}. $E_{b}^{dimer}$ is calculated by 
\begin{equation}
  E_{b}^{dimer} = |E^{dimer} - E_1 - E_2|
  \end{equation}
Here, $E^{dimer}$ is the total energy of the dimer, while $E_1$ and $E_2$ are the energies of the two isolated gas-phase water monomers at equilibrium.
Although PBE yields values close to the CCSD(T) reference results, this apparent agreement should be interpreted with caution. Bare PBE omits vdW attraction but nevertheless overbinds the gas-phase water dimer because of errors in its semilocal description \cite{santra2009CoupledClusterBenchmarks,forster-tonigoldDispersionCorrectedRPBE2014b}. Consequently, adding the D3 correction to PBE does not substantially improve the description of water-dimer binding and may further increase the overbinding. By contrast, revPBE and RPBE underbind the water dimer; When combined with the D3 correction, they yield binding energies and equilibrium O--O distances closer to the CCSD(T) reference values \cite{lin2012StructureDynamicsLiquida}.

\section{Bulk water}
 \begin{table}[htbp]
  \centering
  \caption{Pressure(GPa) and standard deviation within parentheses based on different XC functionals at ambient and 1000 K from the analysis of the stress tensors. The stress tensor was computed using an increased plane wave energy cutoff of 85 Ry by sampling every 2 ps for each trajectory. The values for ambient PBE and SCAN were obtained from the Ref.  \cite{dawson2018equilibration}and \cite{lacount2019ensemble}, respectively.}
    \begin{tabular}{|l|l|l|l|l|l|}
    \hline
        & \multicolumn{1}{r|}{0.88$g/cm^3$} & \multicolumn{1}{r|}{1.32$g/cm^3$} & \multicolumn{1}{r|}{1.44$g/cm^3$} & \multicolumn{1}{r|}{1.57$g/cm^3$} & \multicolumn{1}{r|}{1.00$g/cm^3$*}   \\
    \hline
    PBE   & 1.28(0.6) & 5.50(0.8) & 7.01(0.9) & 10.51(0.9) & 0.54(0.4)  \\
    PBE-D3 & 0.82(0.7) & 4.44(0.8) & 5.78(1.0) & 8.89(1.2) & 0.12(0.4)   \\
    RPBE-D3 & 1.14(0.5) & 4.97(0.9) & 6.40(1.1) & 9.42(1.0) & 0.27(0.4)   \\
    SCAN  & 0.66(0.5) & 3.71(0.7) & 5.06(0.9) & 8.37(0.9) & -0.29(0.4)  \\
    EOS(2009)\cite{zhang2009ModelFluidEarth}& 0.93 &5.04 &6.91 &11.38 & \\
    EOS(2005)\cite{zhang2005PredictionPVTProperties}& 0.91 &4.48 &6.00 &9.5 &   \\
    \hline
    \end{tabular}%
  \label{pressure}
\end{table}%

\begin{figure}[htbp] 
  \centering
  \includegraphics[width=1\textwidth]{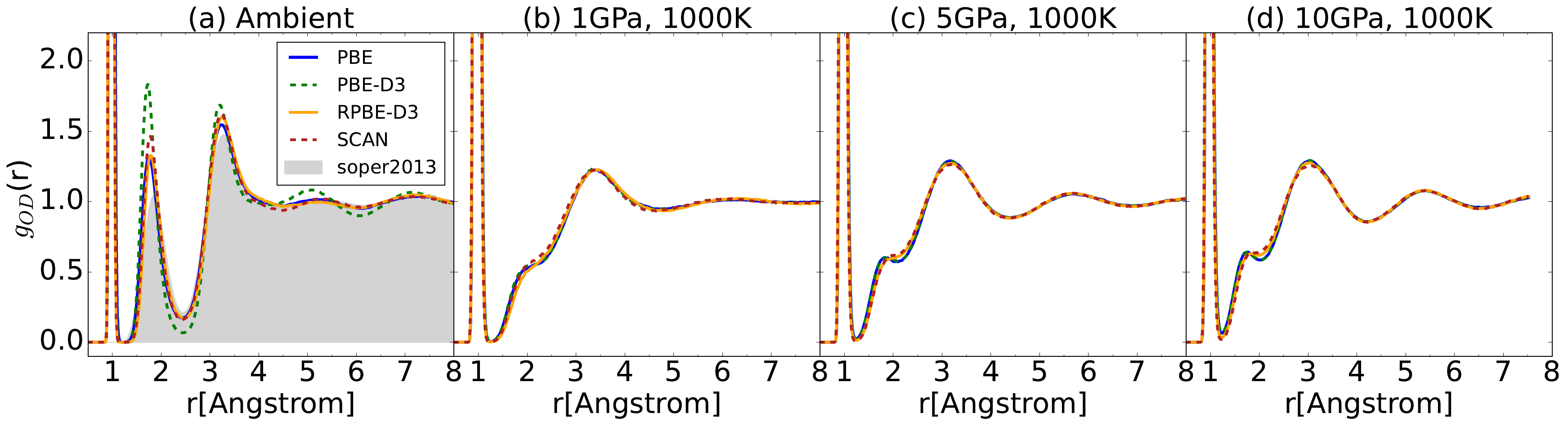}
  \caption{O–D RDFs of water at ambient temperature and 1000 K with a pressure of 1 GPa, 5 GPa and 10 GPa based on the PBE, PBE-D3, RPBE-D3 and SCAN functionals. The results of  PBE(400K) and SCAN(330K) for ambient water are from Ref. \cite{dawson2018equilibration}and \cite{lacount2019ensemble}, respectively}
  \label{o-h}
\end{figure}

\newpage

\begin{figure}[htbp]
  \centering
  \includegraphics[width=1\textwidth]{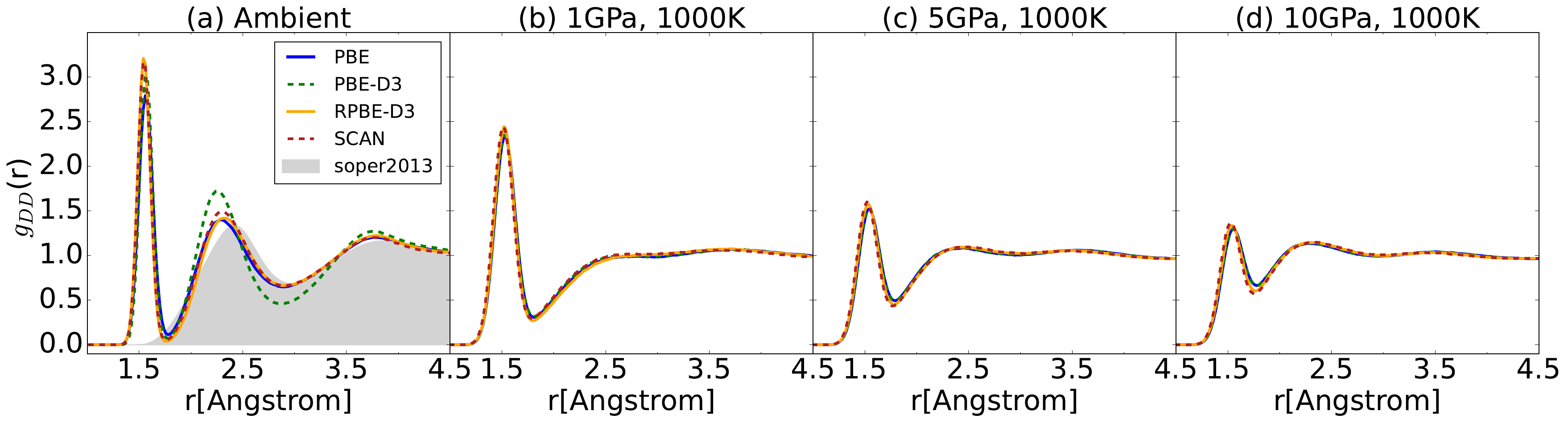}
  \caption{D-D RDFs of water at ambient temperature and 1000 K with a pressure of 1 GPa, 5 GPa and 10 GPa based on the PBE, PBE-D3, RPBE-D3 and SCAN functionals. The results of  PBE(400K) and SCAN(330K) for ambient water are from Ref. \cite{dawson2018equilibration}and \cite{lacount2019ensemble}, respectively}
  \label{h-h}
\end{figure}

\begin{figure}[htbp]
  \centering
  \includegraphics[width=1\textwidth]{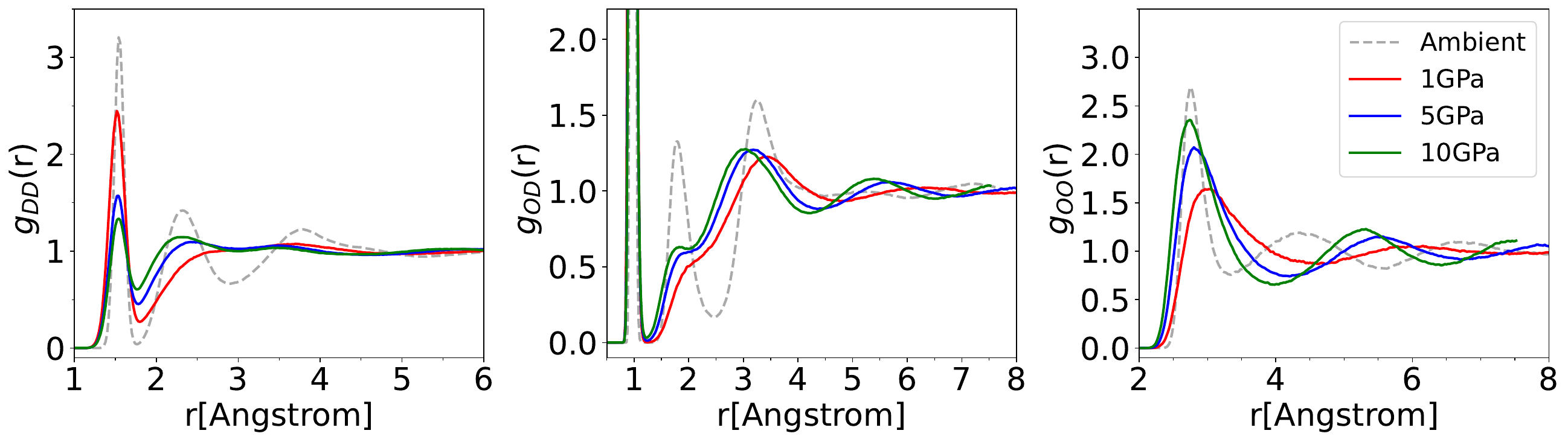}
  \caption{The RDFs given by the RPBE-D3 functional at ambient conditions, and at 1000 K under high pressures.}
  \label{rpbe-d3-rdf}
\end{figure}

\newpage
\bibliography{SI/ref-SI}